# Evaluation and Design Space Exploration of a Time-Division Multiplexed NoC on FPGA for Image Analysis Applications.

Linlin Zhang[1], Virginie Fresse[1], Mohammed. Khalid[2], Dominique Houzet[3], Anne-Claire Legrand[1]

[1] *Université de Lyon, F-42023, Saint-Etienne, France;*
*CNRS, UMR 5516, Laboratoire Hubert Curien, F-42000, Saint-Etienne, France;*
*Université de Saint-Etienne, Jean-Monnet, F-42000, Saint-Etienne, France;*
*{lin.zhang, virginie.fresse, anne.claire.legrand}@univ-st-etienne.fr*
[2] *RCIM: dept. of Electrical & Computer Eng. University of Windsor, Windsor, ON, Canada*
*mkhalid@uwindsor.ca*
[3] *GIPSA-lab, Grenoble, France*
*dominique.houzet@gipsa-lab.grenoble-inp.fr*

*The aim of this paper is to present an adaptable Fat Tree NoC architecture for Field Programmable Gate Array (FPGA) designed for image analysis applications. Traditional NoCs (Network on Chip) are not optimal for dataflow applications with large amount of data. On the opposite, point to point communications are designed from the algorithm requirements but they are expensives in terms of resource and wire. We propose a dedicated communication architecture for image analysis algorithms. This communication mechanism is a generic NoC infrastructure dedicated to dataflow image processing applications, mixing circuit-switching and packet-switching communications. The complete architecture integrates two dedicated communication architectures and reusable IP blocks. Communications are based on the NoC concept to support the high bandwidth required for a large number and type of data. For data communication inside the architecture, an efficient time-division multiplexed (TDM) architecture is proposed. This NoC uses a Fat Tree (FT) topology with Virtual Channels (VC) and flit packet-switching with fixed routes. Two versions of the NoC are presented in this paper. The results of their implementations and their Design Space Exploration (DSE) on Altera StratixII are analyzed and compared with a point to point communication and illustrated with a multispectral image application. Results show that a point-to-point communication scheme is not efficient for large amount of multispectral image data communications. A NoC architecture uses only 10% of the memory blocks required for a point to point architecture but seven times more logic elements. This resource allocation is more adapted to image analysis algorithms as memory elements are a critical point in embedded architectures. A FT NoC-based communication scheme for data transfers provides a more appropriate solution for resource allocation.*

## INTRODUCTION

Image analysis applications consist of extracting some relevant parameters from one or several images or data. Embedded systems for real time image analysis allow computers to take appropriate actions for processing images under real-time hard constraints and often in harsh environments. Current image analysis algorithms are resource intensive so the traditional PC or DSP-based systems are unsuitable as they cannot achieve the required high-performance.

Increases in chip density following Moore's law allow the implementation of ever larger systems on a single chip. Known as systems on chip (SoCs), these systems usually contain several CPUs, memories and custom hardware modules. Such SoCs can also be implemented on FPGA. For embedded real time image processing algorithms, the FPGA devices are widely used because they can achieve high-speed performances in a relatively small footprint with low power compared to GPU architectures M.4[1]. Modern FPGAs integrate many heterogeneous resources on one single chip. The resources on an FPGA continue to increase at a rate that only one FPGA is capable to handle all processing operations, including the acquisition part. That means that incoming data from the sensor or any other acquisition devices are directly processed by the FPGA. No other external resources are required for many applications (some algorithms might use more than one FPGA). Today, many designers of such systems choose to build their designs on Intellectual Property (IP) cores connected to traditional buses. Most IP cores are already pre-designed and pre-tested and they can be immediately reused M.4 M.4 M.4. Without reinventing the wheel, the existing IPs and buses are directly used and mapped to build the dedicated architecture. Although the benefits of using existing IPs are substantial, buses are now replaced by NoC communication architectures for a more systematic, predictive and reliable architecture design. Network on Chip architectures are classsified according to their switching technique and to their topology. . Few NoC architectures for FPGA are proposed in the literature. Packet switching with wormhole is used by Hermes M.4, IMEC M.4, SoCIN M.4, and Extended Mesh

M.4 NoCs. PNoC M.4 and RMBoC M.4 use only the circuit switching whereas the NoC of Lee H.G M.4 uses the packet switching. For the topology, Hermes uses a 2D mesh, the NoC from IMEC uses a 2D torus, SoCIN/RASoC can use a 2D mesh or a torus. RMBoC from M.4 has a 1D or 2D mesh topology. An extended mesh is used for the Extended Mesh NoC. HIBI uses a hierarchical bus. PNoC and the NoC from Lee H.G have a custom topology.
Existing NoC architectures for FPGA are not adapted to image analysis algorithms as the number of input data is high compared to the results and commands. A dedicated and optimized communication architecture is required and is most of the time designed from the algorithm requirements.
The Design Space Exploration (DSE) of an adaptable architecture for image analysis applications on FPGA with IP designs remains a difficult task. It is hard to predict the number and the type of the required IPs and buses from a set of existing IPs from a library.

In this paper we present an adaptable communication architecture dedicated to image analysis applications. The architecture is based on a set of locally synchronous modules. The communication architecture is a double NoC architecture, one NoC structure dedicated to commands and results, the other one dedicated to internal data transfers. The data communication structure is designed to be adapted to the application requirements (number of tasks, required connections, size of transmitted data). Proposing a NoC paradigm helps the dimensioning and exploration of the communication between IPs as well as their integration in the final system.

The paper is organised into 5 further sections. Section 1 presents the global image analysis architecture and focuses on the data flow. Special communication units are set up to satisfy the application constraints. Section 2 presents two versions of NoC for the data flow which are built on these basic communication units. The NoC architectures are totally parameterized. Section 3 presents one image analysis application : a multispectral image authentication. DSE method is used to find out the best parameters for the NoC architecture according to the application. Section 6 gives the conclusion and perspectives.

## Architecture dedicated to Image Analysis algorithms

This architecture is designed for most of image analysis applications. Characteristics from such applications are used to propose a parameterized and adaptable architecture for FPGA.

### A Characteristics of image analysis algorithms

Image analysis consists of extracting some relevant parameters from one or several images. Image analysis examples are object segmentation, feature extraction, motion detection and object tracking, etc M.4 M.4. Any image analysis application requires four types of operations:

- Acquisition operations.
- Storage operations.
- Processing operations.
- Control operations.

A characteristic of image analysis applications is the unbalanced data flow between the input and the output. The input data flow corresponds to a high number of pixels (images) whereas the output data flow represents little data information (selective results). From these unbalanced flows, two different communication topologies can be defined, each one being adapted to the speed and flow of data.

### B An adaptable structure for image analysis algorithms

The architecture presented here is designed from the characteristics of image analysis applications. The structure of the architecture contains four types of modules, each one corresponds to the four types of operations. All these modules are designed as several VHDL Intellectual Property (IP) nodes. They are presented in details in M.4:

- The **Acquisition Module** produces data that are processed by the system. The number of acquisition modules depends on the applications and the number of types of required external interfaces.
- The **Storage Module** stores incoming images or any other data inside the architecture. Writing and reading cycles are supervised by the control module. Whenever possible, memory banks are FPGA-embedded memories.
- The **Processing Module** contains the logic that is required to execute one task of the algorithm. The number of processing modules depends on the number of tasks of the application. Moreover, more than one identical processing module can be used in parallel to improve timing performances.
  The number of these modules is only limited by the size of the target FPGA. The control of the system is not distributed in all modules but it is fully centralized in a single control module:

- The **Control Module** performs decisions and scheduling of operations and sends commands to the other modules. All the commands are sent from this module to the other modules. In the same way, this module receives result data from the processing modules.

The immediate reuse of all modules is possible as all modules are designed with an identical structure and interface given in the Fig 1.

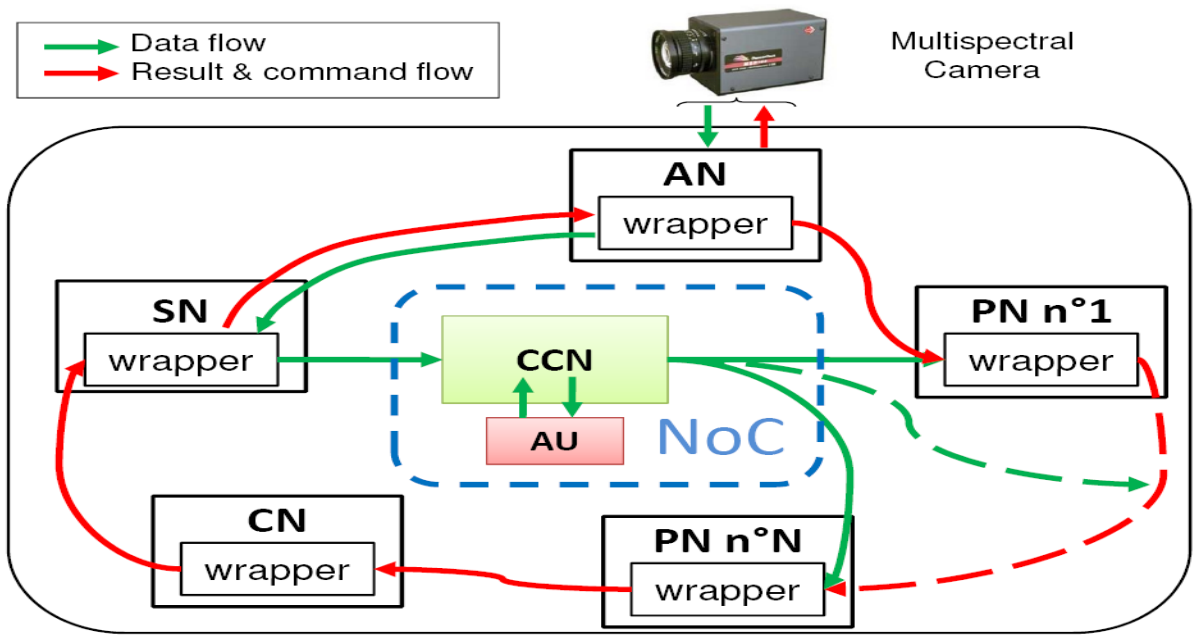

**Fig 1 The proposed adaptable architecture dedicated to image analysis applications.**

To run each node at its best frequency, Globally Asynchronous Locally Synchronous (GALS) concept is used in this architecture. The frequencies for each type of nodes in the architecture depend on the system requirements and tasks of the application.

## C Structure of modules/ NoC for command and results

The modular principle of the architecture can be shown at different levels: one type of operation is implemented by means of a module (acquisition, storage, processing, etc). Each module includes units that carry out a function (decoding, control, correlation, data interface, etc), and these units are shaped into basic blocks (memory, comparator, etc). Some units can be found inside different modules. Fig 2 depicts all levels inside a module.

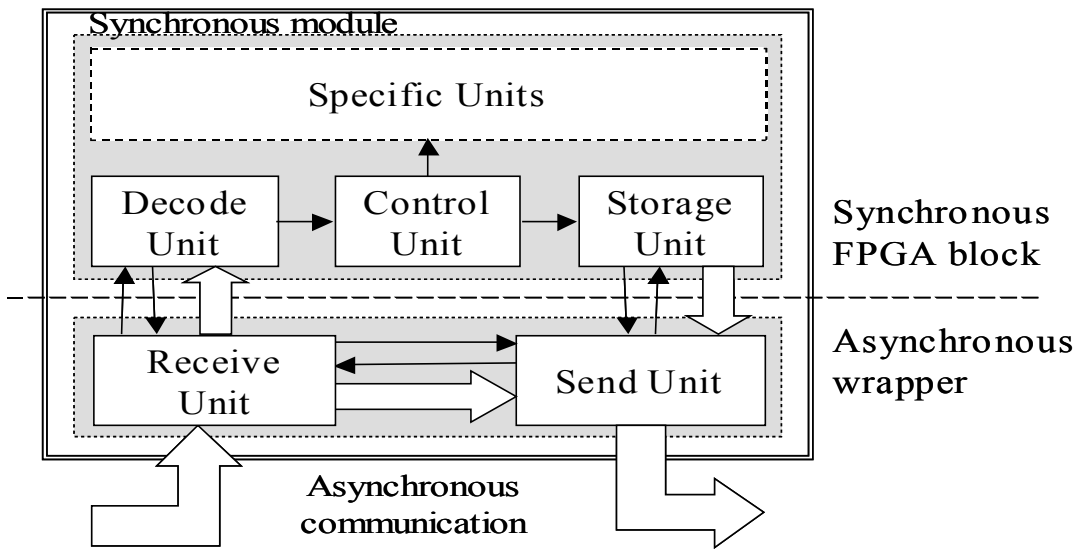

**Fig 2 The generic structure of modules with the asynchronous wrapper for result and command.**

Each module is designed in a synchronous way having its own frequency. Communications between modules are asynchronous via a wrapper and use a single-rail data path 4-phase handshake. Two serial flip-flops are used between independent clock domains to reduce the metastability M.4 M.4. The wrapper includes two independent units. One receives frames from the previous module and the other one sends frames to the following module at the same time.

A NoC is characterized by its topology, routing protocol and flow control. The communication architecture is a NoC for command and results and another NoC for internal data. Topology, flow control and type of packets differ according to the targeted NoC.

## D NoC for command and results

Because the command flow and the final results are significantly fewer compared to the incoming data, they use a NoC architecture which is linked to the IP wrappers. The topology for this communication is a ring using a circuit switching technique with 8 bit flits. Through the communication ring, the control module sends 4 packets. Packets have one header flit and 3 other flits containing command flits and empty flits. The control module sends packets to any other modules, packets are command packets or empty packets. Command packets sent by the control module to any other module contain instructions to execute. Empty packets are used by any other module to send data to the control module. Empty packets can be used by any module to send results or any information back to the control module.

# E Communication architecture for data transfers

The NoC dedicated to data uses a Fat Tree topology which can be customized according to the required communication of the application. Here we use *flit packet-switching/wormhole routing with fixed routes and virtual channels*. Flow control deals with the allocation of channel and buffer resources to a packet/data. For image analysis applications, the specifications for the design of our NoC dedicated to data are:

- Several types of data with different lengths at the inputs. The size of the data must be parameterized to support any algorithm characteristic.
- Several output nodes, this number is defined according to the application requirements.
- Frequencies nodes/modules are different.

According to the algorithms implemented, several data from any input module can be sent to any output module at any time.

In the following sections, we assume that the architecture contains four input modules (the memory modules) connected to four output modules (the processing modules). This configuration will be used for the multispectral image application illustrating the design space exploration in the following sections.

## E.1. The topology

The topology chosen is a Fat Tree (FT) topology as depicted in Fig. 3 as it can be adapted to the algorithm requirements. Custom routers are used to interconnect modules in this topology.

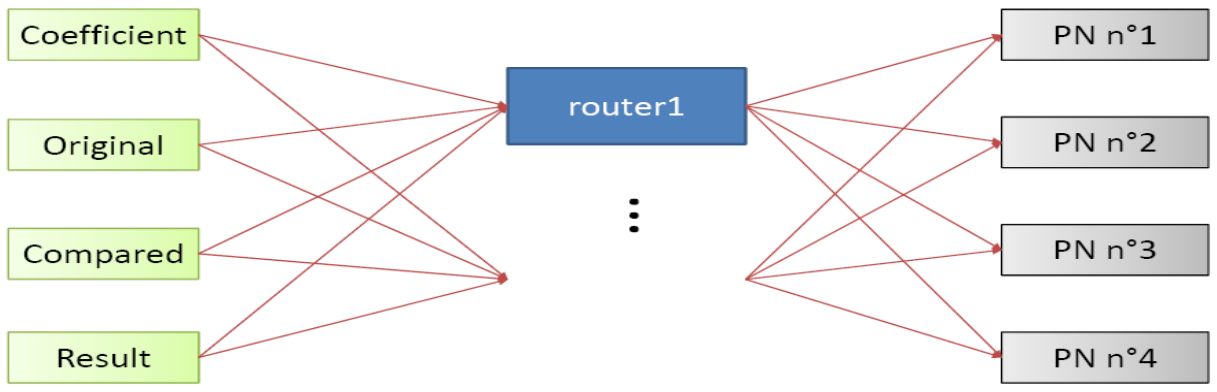

**Fig 3 FT topology for the TDM NoC.**

## E.2. Virtual Channel (VC) flow control

VC flow control is a well-known technique. A VC consists of a buffer that can hold one or more flits of a packet and associated state information. Several virtual channels may share the bandwidth of a single physical channel M.4. It allows minimization of the size of the router's buffers – a significant source of area and energy overhead M.4 M.4, while providing flexibility and good channel use.

During the operation of the router a VC can be in one of the states: idle, busy, empty or ready.

Virtual channels are implemented using bi-synchronousd FIFOs.

## E.3. Packet/ Flit structure

| Header 8 bit | | | | | | | | Data N bit [0,9].[0,9][0,9][0,9] | | | | | | | | | | | | | | | | Tail 8 bit | | | |
|---|---|---|---|---|---|---|---|---|---|---|---|---|---|---|---|---|---|---|---|---|---|---|---|---|---|---|---|
| id | | p | | | Int length | | | 1st integer | | | | 2nd integer | | | | 1st floating | | | | Nth floating | | | | Constants | | | |
| | | | | | | | | | | | | | | | | | | | | | | | | F | | F | |
| . | . | . | . | . | . | . | . | . | . | . | . | . | . | . | . | . | . | . | . | 12 | 11 | 10 | 9 | 8 | 4 | 3 | 0 |

**Fig 4 Data structures for the packets.**

Fig 4 shows the structure of the packet/flit used for the data transfers. The packet uses a 8-bit header flit, a 8-bit tail flit and several 8-bit flits for the data. For the header flit, Id is the IDentified number of data. P is the output Port number corresponding to the number of PN. Int_l signifies INTeger Length represents the position of the fixed point in data.

The *tail* flit is a constant "FF". One packet can be separated in several Flow control unit (flit). The data structure is dynamic in order to adapt to different types of data. The length of packet and data, number and size of flits and the depth of VC are all parameterized. The size of flits can be 8, 16, 32 or 64 bits, but we keep a header and tail of 8 bits, extended to the flit size.

Packet switching with wormhole and fixed routing paths is used, each packet containing the target address information as well as data with Best Effort (BE) traffic.

## E.4. The switch structure

This NoC is based on routers built with three blocks. One block called Central Coordination Node (CCN) performs the coordination of the system. The second block is the Arbitration Unit (AU) which detects the states of data paths. The last one is a mux (TDM-NA) with formatting of data. The switch structure is shown in Fig 5.

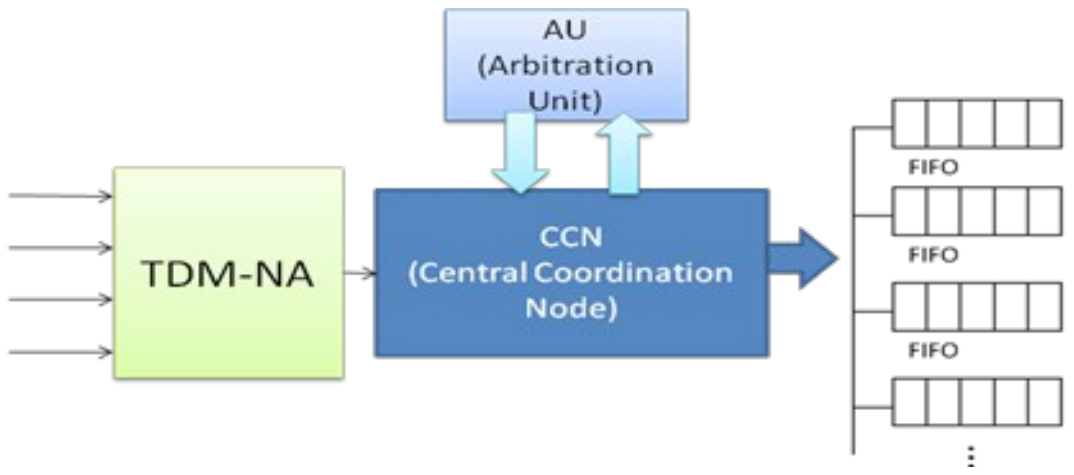

**Fig 5 Switch structure.**

The CCN manages the resources of the system and maps all new incoming messages to the target channel. The switch is based on a mux (crossbar) from several inputs to several outputs. All the inputs are multiplexed using the TDM Time Division Multiplexing.

For a high throughput, more than one switch can be implemented in the communication architecture.

The AU is a Round Robin Arbiter (RRA) M.4 M.4 which detects the states of all the VC at the outputs. It determines on a cycle-by-cycle basis which VC may advance. When AU receives the destination information of the flit (P_enc), it detects the available paths' states connected to the target output. This routing condition information will be sent back to CCN in order to let CCN perform the mapping of the communication.

## E.5. The structure of TDM-NA

The TDM-NA is a set of a MUX and a Network Adapter (NA). One specific NA is proposed in Fig 6. The Network Adapter adapts any data before being sent to the communication architecture. The Network Adapter contains 5 blocks.

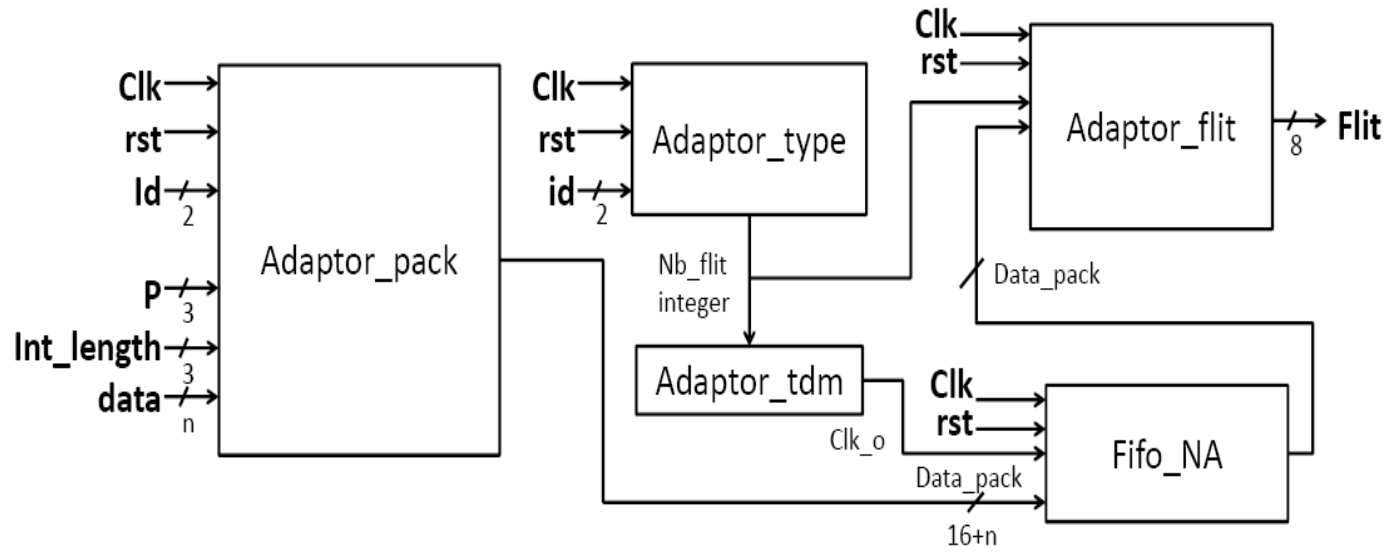

**Fig 6 Data structure for the defined types of data.**

- Adaptor_type: verifies the type of the data and defines the required number of flits (and the number of clock cycles used to construct the flits).
- Adaptor_tmd: performs the time division multiplexing for the process of cutting one packet to several flits.
- Adaptor_pack: add *header* and *Tail* flits to the initial data.
- Fifo_NA: stores the completed packet.
- Adaptor_flit: cuts the whole packet into several 8-bit flits.

Adaptor_flit runs with a higher clock frequency in this NA architecture because it needs time to cut one packet into several flits. For different lengths of data, Adaptor_tdm will generate different frequencies which depend on the number of flits going out for one completed packet..

## Two Versions of the TDM Parameterized NoCs

Two versions are proposed and presented in this paper. Data are transferred in packets in version 1 with a packet switching technique and with a fixed size of links. Data are transferred with flits with a wormhole technique and a reduced size of the links in version 2. The first version uses one main switch and 2 Virtual Channels on the outputs. The second version contains 2 main switchs in parallel with 2 Virtual Channels on the inputs and on the outputs. All versions have four memory modules as input modules and four processing modules as output modules. All versions are designed in VHDL.

## F Version 1 with ONE main switch

Version 1 is a TDM FT NoC containing one main switch and 2 VCs, 2 channels for each output as shown in Fig 7.

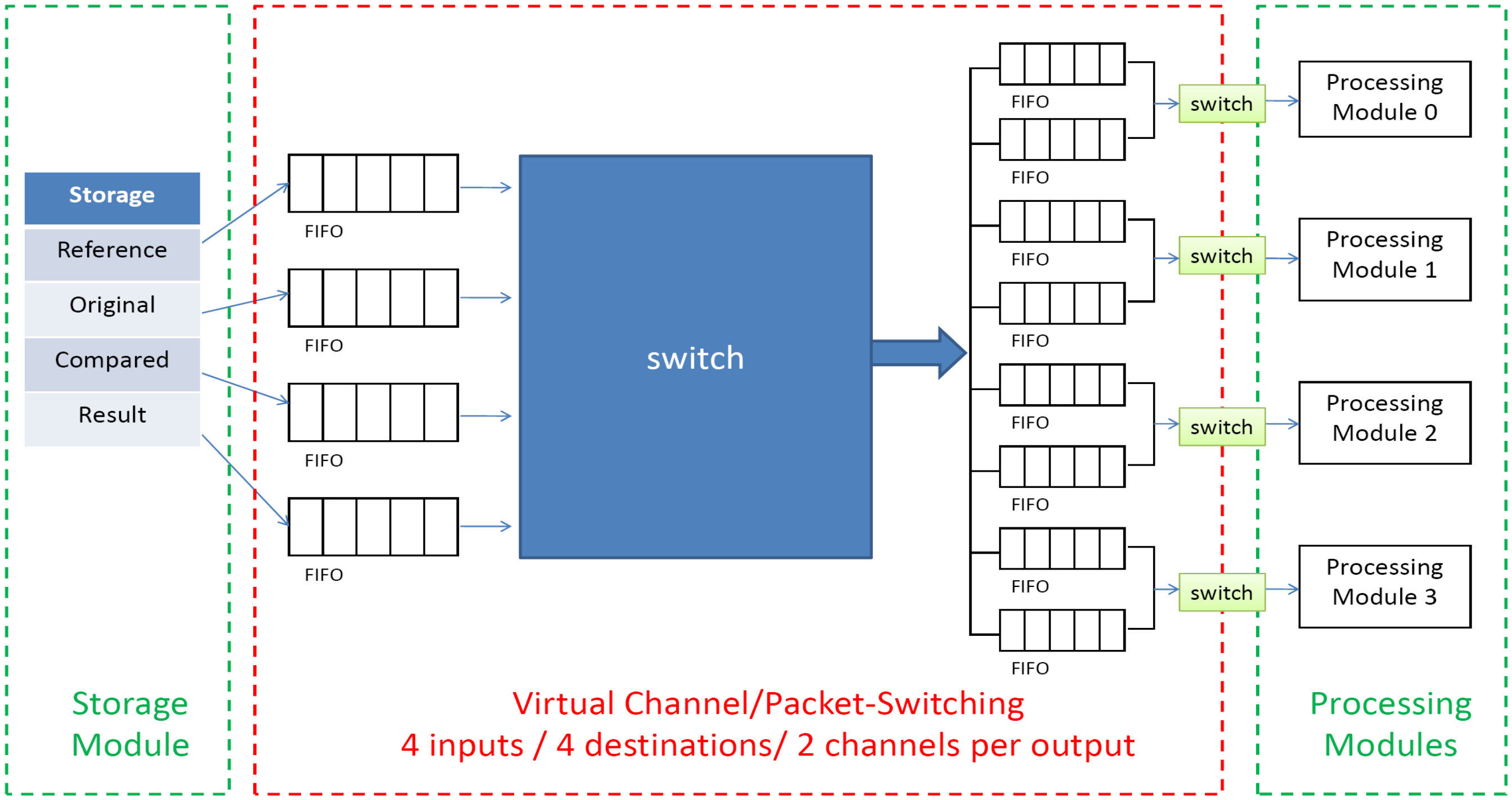

**Fig 7 The structure of Version 1**

the data are sent as 24-bit packets. The width of VCs in version 1 is 24 bits. The simplified data structure of version 1 is shown in Fig 8.

| Head 8 bit | | | | | | | | Data 16 bit [0,15].[0,10][0,10][0,10] | | | | | | | | | | | | | | | |
|---|---|---|---|---|---|---|---|---|---|---|---|---|---|---|---|---|---|---|---|---|---|---|---|
| id | | p | | | No used | | | Integer | | | | 1st floating | | | | 2nd floating | | | | 3rd floating | | | |
| | | | | | | | | | | | | | | | | | | | | | | | |
| 27 | 26 | 25 | 24 | 23 | 22 | 21 | 20 | 19 | 18 | 17 | 16 | 15 | 14 | 13 | 12 | 11 | 10 | 9 | 8 | 7 | 6 | 5 | 4 |

**Fig 8 The 24 bit packet data structure for Version 1**

## G Version 2 with TWO main switches

Another switch is added to the architecture to increase the throughput. Structure of switch is identical to the switch presented in the previous section. These two main switches operate in parallel as depicted in Fig 9.

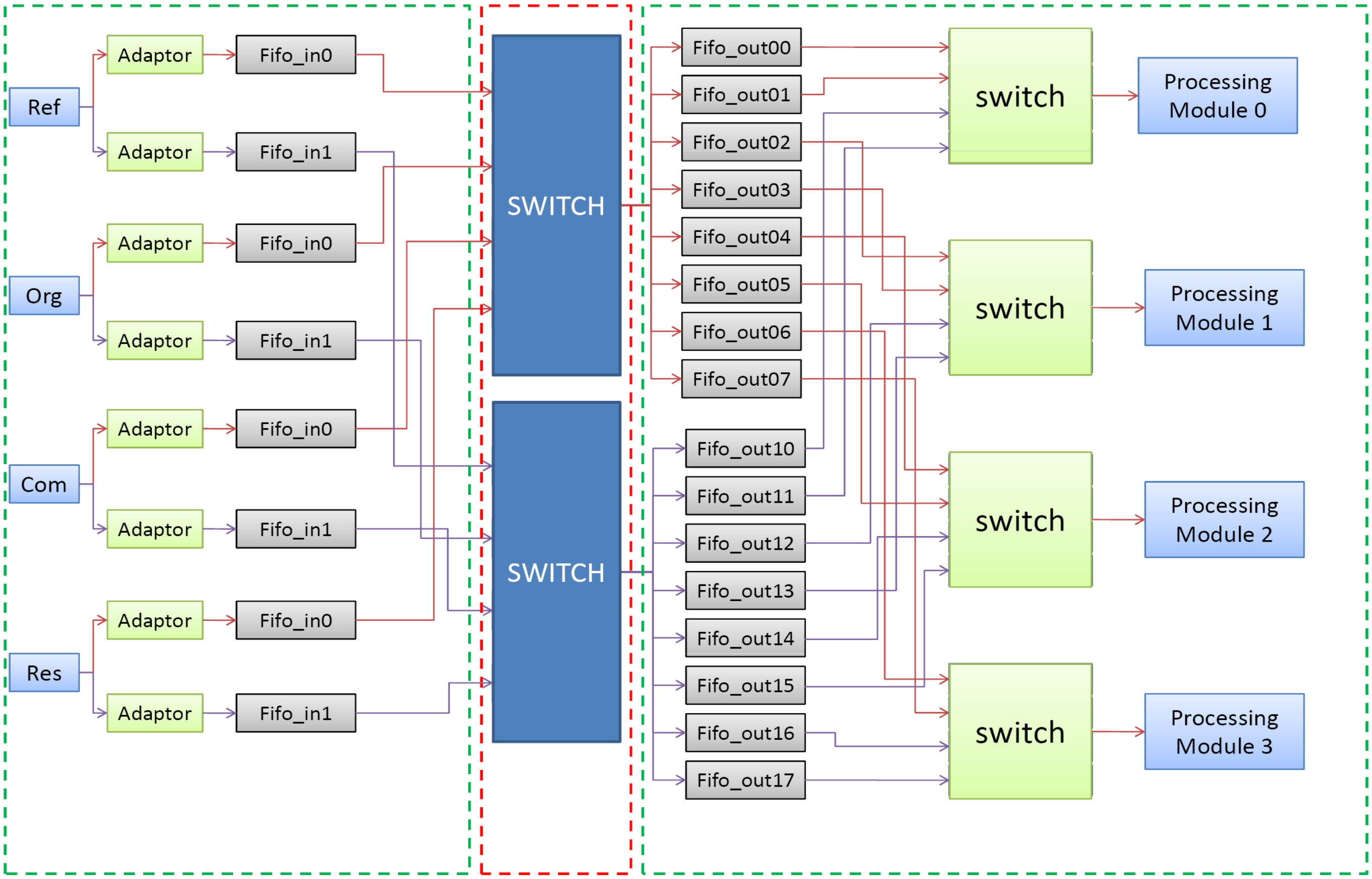

**Fig 9 The structure of version 2 with 2 main switches in parallel.**

The width of all the VCs in this version depends on the algorithm characteristics..

## H NoC Parameters for DSE

The proposed NoC is flexible and adaptable to the requirements of image analysis applications. Parameters from the communication architecture are specified for the Design Space Exploration (DSE).
The parameters are:

- Number of switches: one main switch for version 1 and two main switches for version 2.
- Size of VCs: it corresponds to the different sizes of the different types of data transferred.
- Depth of the FIFOs in VCs: limited by the maximum storage resources of the FPGA.

Several synthesis tools are used for the architecture implementation and DSE as these synthesis tools give different resource allocations on FPGA.

## Experiments and Results

The size of data, FIFOs and virtual channels are extracted from the algorithm implemented. A multispectral image algorithm for image authentication is used here to validate the communication architecture.

## I Multispectral Image Authentication

Multispectral image Fig 10analysis(Fig 10) has been used in the space-based image identifications since 1970s M.4 M.4 M.4 M.4 M.4. This technology can capture light from a wide range of frequencies. This can allow extraction of additional information that the human eye fails to capture with its receptors of red, green and blue. Art authentication is one common application widely used in museums. In this field, an embedded authentication system is required.

The multispectral images are optically acquired in more than one spectral or wavelength interval. Each individual image has usually the same physical area and scale but a different spectral band. Other applications are presented in M.4 M.4.

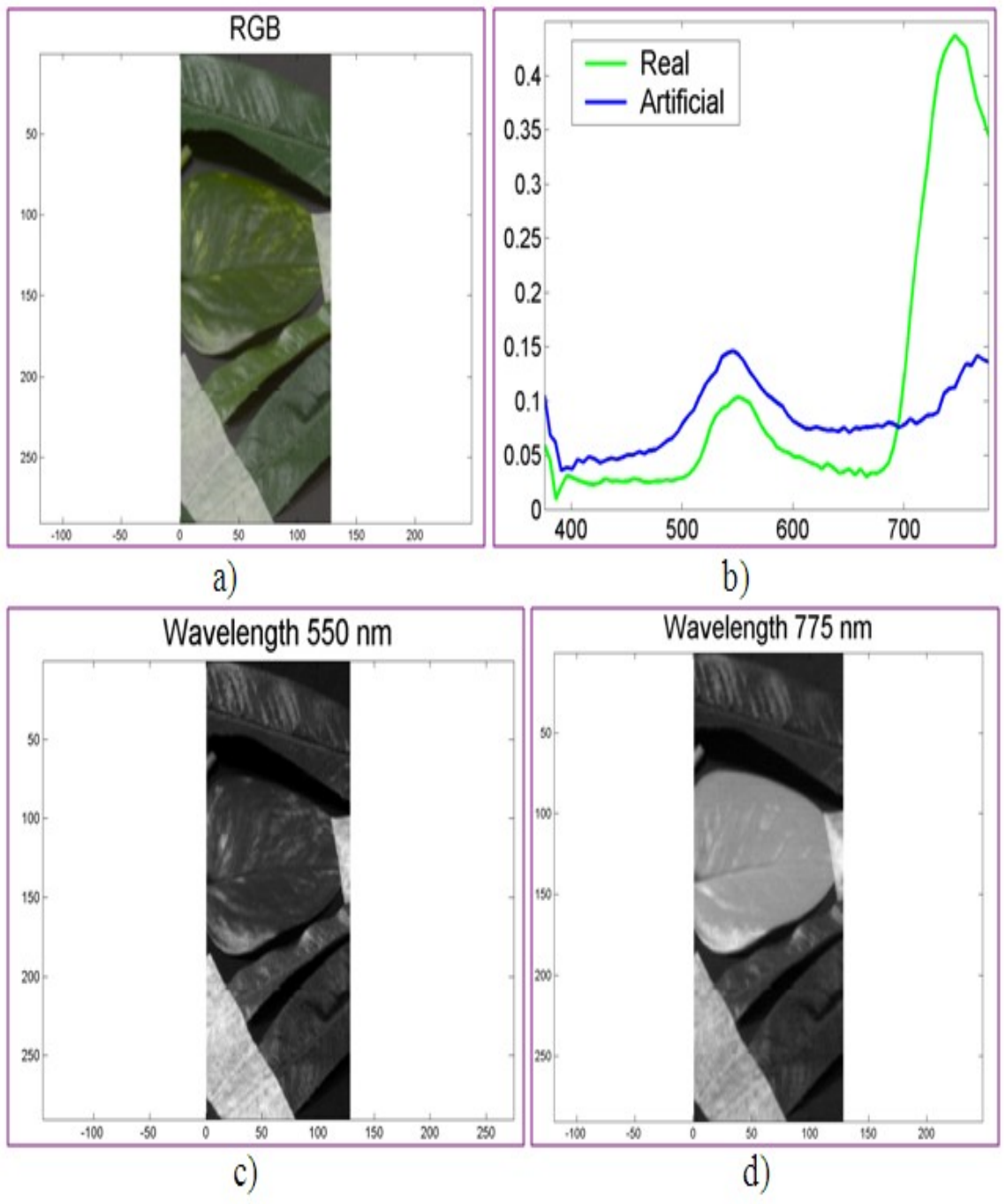

**Fig 10 Multispectral images . Multispectral images add additionnal information compared to color image. In this example, the artificial leaf can be extracted. To the real ones for 775 nm**

The aim of the multispectral image correlation is to compare two spectral images:

- Original image (OI): its spectrum is saved in the Storage Module as the reference data.
- Compared images (CI): its spectrum is acquired by a multispectral camera.

For the art authentication process, OI is the information of the true picture, and the CI are the others "similar" candidates. With the comparison process of the authentication (Fig 11), the true picture can be found among the false ones by calculating the distance of the multispectral image data. For this process, certain algorithms require high precision operations which imply large amount of different types of data (e.g. floating-point, fixed-point, integer, BCD encoding, etc) and complex functions (e.g. square root or other nonlinear functions). Several spectral projections and distance algorithms can be used in the multispectral authentication.

We can detail the process:

- First of all, the original data received from the multispectral camera are the spectral values for every pixel on the image. The whole image will be separated as several significant image regions. These regions' values need to be transformed as average color values by using different windows' sizes (e.g. 8×8 pixel as the smallest window, 64×64 pixel as the biggest window).
- After this process, certain "color projection" (e.g. RGB, L*a*b*, XYZ, etc) will transform the average color values to color space values. An example of RGB color projection is shown in Equation (1):

$$R_i = \sum_{\lambda=380}^{780} S(\lambda) \times R_c(\lambda)$$

$$G_i = \sum_{\lambda=380}^{780} S(\lambda) \times G_c(\lambda) \qquad (1)$$

$$B_i = \sum_{\lambda=380}^{780} S(\lambda) \times B_c(\lambda)$$

Rc, Gc, and Bc are the coefficients of the Red, Green, Blue color space. S(λ) represents the spectral value of the image corresponding to each scanned wavelength λ. The multispectral camera used can scan one picture from 380nm to 780nm with 0.5nm as precision unit. So the number of spectral values N can vary from 3 to 800. Ri, Gi, and Bi are the RGB values of the processed image.

- These color image data go through the comparison process of the authentication. Color distance is just the basic neutral geometry distance. For example, for the RGB color space, the calculated distance is shown in Equation 2:

$$\Delta E_{RGB} = \sqrt{(R_1 - R_2)^2 + (G_1 - G_2)^2 + (B_1 - B_2)^2} \quad (2)$$

If the true picture can be found among the false ones by calculating the color distance, the process is finished otherwise goes to the next step.

- Several multispectral algorithms (e.g. GFC, Mv) are used to calculate the multispectral distance with the original multispectral image data. Certain algorithms require high precision operations which imply large amount of floating-point data and complex functions (e.g. square root or other nonlinear functions) in this process.
- After comparing all the significant regions on the image, a ratio (Rs/d) of similitude will be calculated as shown in Equation (3). N°s represents the number of similar regions and N°d represents the number of dissimilar regions.

$$R_{s/d} = N^{\circ}_{s} / N^{\circ}_{d} \quad (3)$$

Different thresholds will be defined to give the final authentication result for the different required precisions: finding the true image which is most alike the original one. One of these algorithms is presented in M.4.

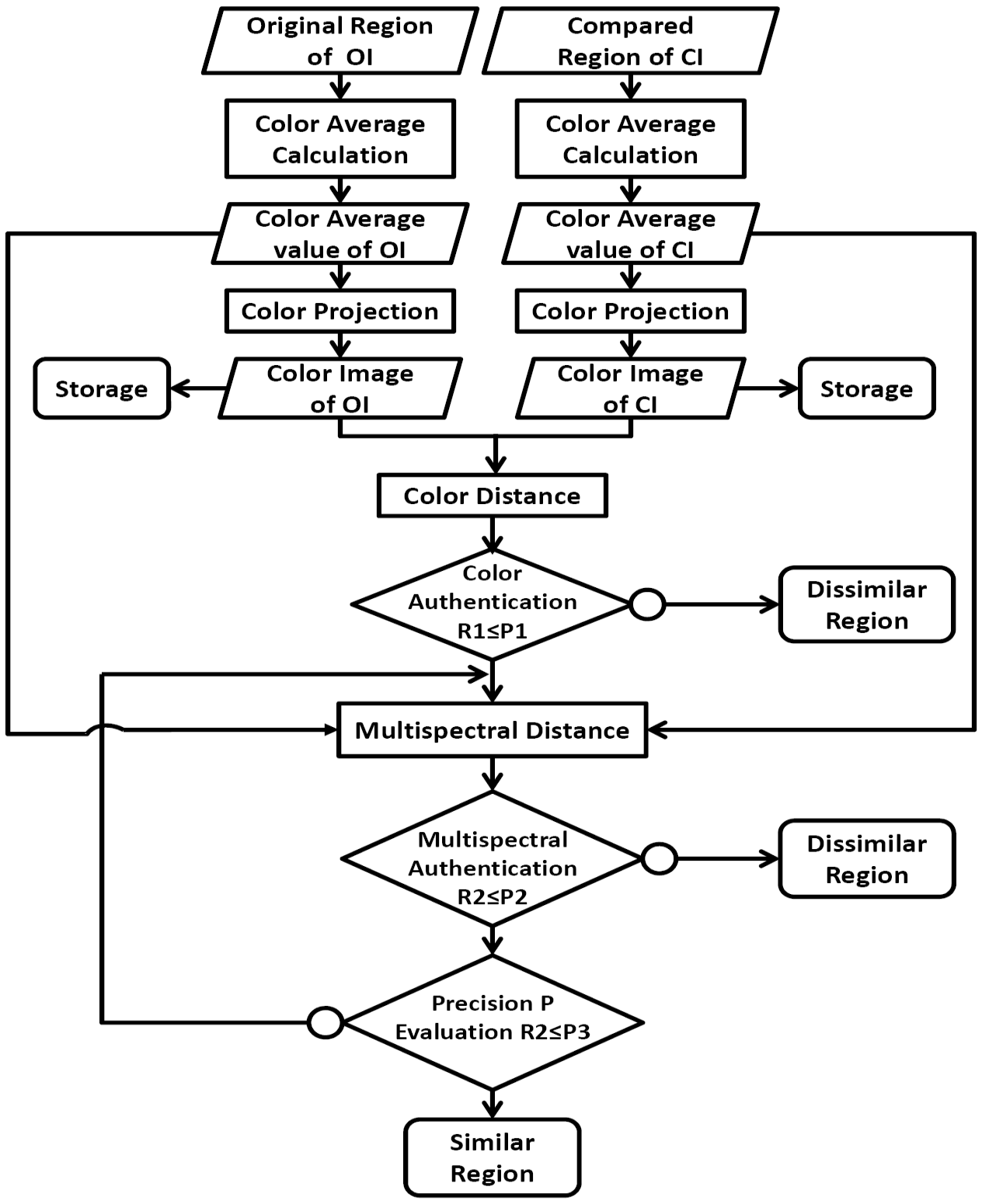

**Fig 11 General comparison process of the authentication**

**R–Result of each step of calculation**
**P–Precision of each multispectral distance**

The calculations are based on the spatial and spectral data which make the memory accesses a bottleneck in the communication. From the algorithm given in Fig. 11, the characteristics are:

■Number of regions for every wavelength = 2000
■Number of wavelength= 992
■Size of the window for the average processing= 2x2, 4x4, 8x8, 16x16, 32x32.
■Number of tasks: 4. color projection, color distance, multispectral projection, multispectral distance. The multispectral authentication task is executed by the control module. In this example, there is no task parallelism. Sizes of data are 72 bits, 64 bits, 64 bits and 24 bits as shown in Fig 12.
■Number of modules: 4 processing modules, 4 storage modules, 1 acquisition module, 1 control module.
■Bandwidth of multispectral camera: 300 MB/s. The NoC architecture is dimensioned to process and exchange data at least at the same rate in order to achieve real-time.

For the NoC architecture, four types of data are defined by analyzing multispectral image algorithms. Each data has an identical number *id)*:

- Coef: Coefficient data which means the normalized values of difference color space vector. (56-bit, *id "00"*).
- Org: Original image data which are stored in the SN. (48-bit, *id "01"*).
- Com: Compared image data which are acquired by the multispectral camera and received from the NA. (48-bit, *id "10"*).
- Res: Result of the authentication process. (8-bit, *id "11"*).

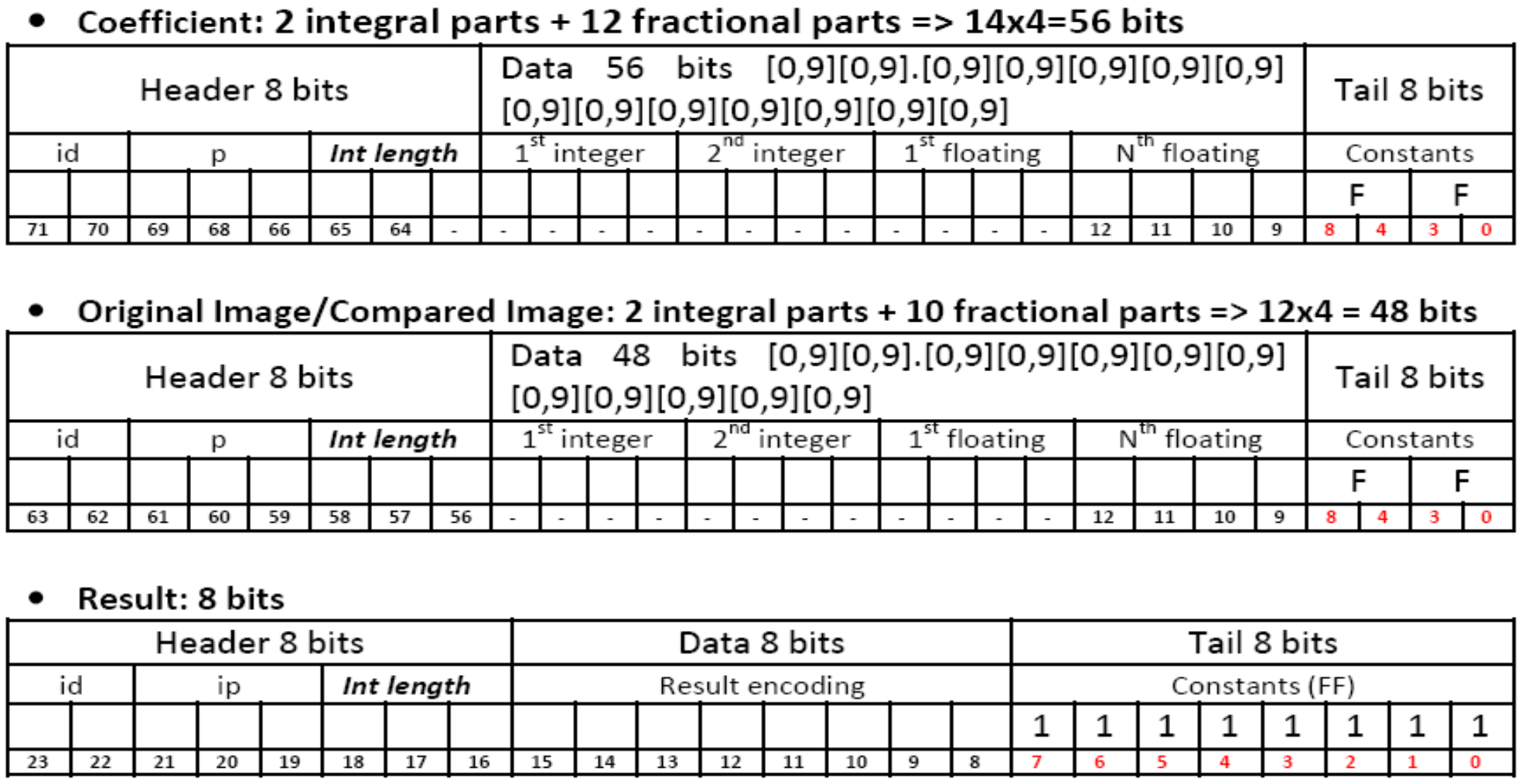

**Fig 12 type and size of data for the multispectral algorithm.**

## J Resources of modules in the architecture

This parameterized TDM architecture was designed in VHDL. Table 1 shows the resources of the modules in the architecture. The FPGA is the Altera Stratix II EP2S15F484C3 which has 6240 ALMs/logic cells. The number of resources dedicated to all the modules represent around 14% of the total logic cells. Whatever the communication architecture, all these modules remain unchanged with the same number of resources.

**Table 1 Resources for the Nodes in the GALS Architecture**

| Node | Frequency (MHz) | Resources on StratixII 2S60 | | |
|---|---|---|---|---|
| | | Logic cells | Registers | Memory bits |
| Control | 150 | 278 | 265 | 32 |
| Acquisition | 76.923 | 315 | 226 | 2 |
| Storage | 100 | 280 | 424 | 320000 |
| Processing | 50 | Depending on the algorithms | | |

## K The point to point communication architecture dedicated to multispectral image algorithms

A classical point-to-point communication architecture is designed for the algorithm requirements presented previously and is shown in Fig 13. This traditional structure is used to compare some significant results obtained by the proposed NoC. In the global communication architecture, any input data can be transmitted to any processing module. 72 bit-muxes are inserted here between FIFO and processing modules. This point-to-point communication uses input FIFOs having the size of data used. Their bandwidth is thus not tuned to fit the bandwidth of the input streams. For the three versions studied here, the input FIFO bandwidth is higher than the specifications of multispectral cameras. If it was not the case, the input FIFO size could be increased to respect the constraint.

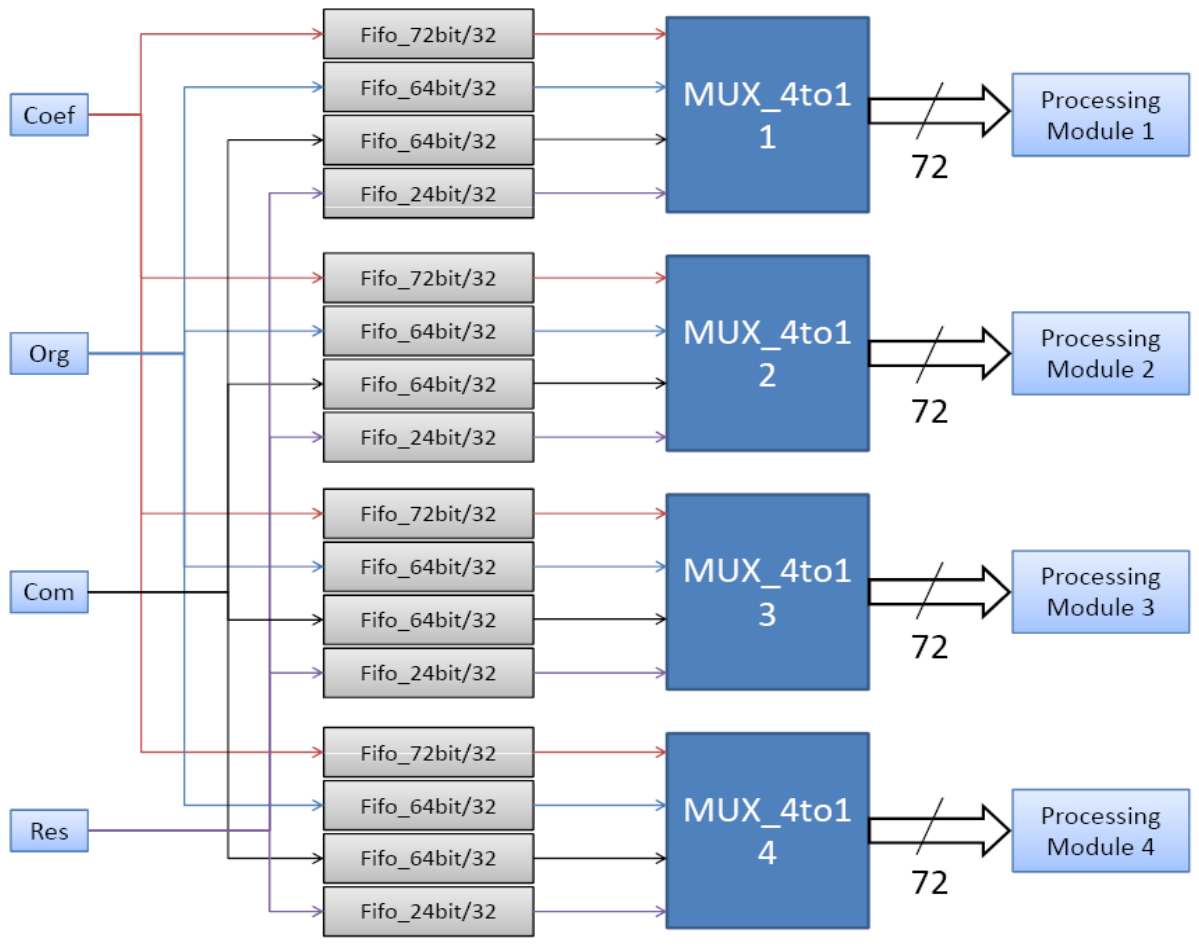

**Fig 13 The point-to-point communication dedicated to the multispectral authentication algorithm.**

## L Implementation of the communication architecture for data transfers

The point to point architecture and both versions of the NoC are designed in VHDL language. The FPGA used for the implementation is an Altera StratixII EP2S15F484C3 EP2S180F1508C3 FPGA. Two sizes of data are used for version 1 of the NoC, 48-bits and 56-bits. These sizes are similar to the size of data for the point to point communication. Implementation results are given in Table 2.

Concerning latency, version 2 uses 8-bit flits as the transmission unit, thus the NA needs 8 cycles to cut a 64-bit data packet as flits plus 1 cycle for the header. Also, the latency of the NoC is 1 cycle for the storing in the first FIFO, 1 cycle for the main switch crossing and 1 cycle for the storing in the second FIFO, that is 3 cycles of latency due to the NoC. Compared to the point-to-point communication, we pay the packet serialization latency to have much better flexibility.

**Table 2 Comparison of the resources: Point-to-point VS. NoC Version 1**

| Ressources | Point-to-Point | Version 1 48-bit | Version 1 56-bit | Version 2 |
|---|---|---|---|---|
| Logic Utilization % | 1% | 2% | 3% | 3% |
| Combinational ALUTs | 305 | 1842 | 2118 | 2521 |
| Dedicated logic registers | 1425 | 2347 | 2739 | 4217 |
| Total pins | 512 | 344 | 408 | 230 |
| Total block memory bit | 29568 | 3384 | 3960 | 8652 |
| Frequency for F (MHz) | 165.73MHz | 264.34MHz | 282.41MHz | 292.31MHz |

Concerning area resources, as depicted in Table 2, the point-to-point communication needs less ALUTs but over 7 times more memory blocks. The switch requires 4 times more logic (ALUTs) than the point to point architecture (the other ALUTs are for the FIFOs of versions 1 and 2 that use more registers than memory blocks to implement FIFOs). One reason is the structure of the switch which is more complex than muxes used in the point to point architecture. If we compare just the switch size with a simple classical NoC like Hermes, we obtain similar sizes for a switch based on a 4x4 crossbar, but a full NoC linking 4 memory nodes to 4 processing nodes would require at least 8 switches, that is almost 8 times more area and from 2 times to 5 times more latency to cross the switches when there is no contention and even more with contentions. The advantage of a classical NoC approach is to allow any communication. This is of no use here as our four memories are not communicating together. We have here an oriented dataflow application with specific communications. Our dataflow NoC has the advantages of NoC, that is systematic design, predictability and so on, and the advantages of point to point communications, that is low latency and optimized well sized links to obtain the best performance/cost trade-off and to use less memory blocks which is important for image algorithms using huge quantity of data to be stored inside the chip.

Also the number of pins for the point to point communication is significantly higher compared to both NoC versions, even with a simple communication pattern. It indicates that the point to point communication requires much more wires inside the chip when the complete system is implemented. This can be a limitation for complex Multiprocessor SoC. Furthermore, the frequency of point-to-point communication is a bit slower than NoC versions.

Resource allocations show the benefits of using NoC architecture for image analysis algorithms compared to traditional point to point architectures. The following implementations focus on the NoC architectures. Both versions are analyzed and explored when implemented on FPGA.

The number of resources for the version 1 is less than for the version 2, but with less bandwidth (a single switch compared to 2 switches for version 2). The choice of one version of the NoC is made from the tradeoff on timing and resources. The optimization of the number of resources leads to the choice of version 1 whereas version 2 is adapted to higher bandwith requirements.

# M DSE of the proposed NoC

The knowledge of the design space of architectures is important for the designer to make trade-offs related to the design. The architecture most adapted to the algorithm requirements requires a Design Space Exploration (DSE of the NoC architectures). The exploration presents the relationships between input parameter values and performance results in order to make trade-offs in the implemented architecture. In DSE the parameter values guide the designer to make choices and trade-offs while satisfying the required performances without requiring implementation processes:

- Input parameters:
  - Number of switches
  - Number and width of VCs
  - Depth of FIFOs/VCs
- Performances:
  - Logic device
  - ALUTs
  - Registers
  - Latency
  - Frequency.

The input parameters are explored to see their effect on the performances. Performances are focused on the resources first. The purpose of DSE is to find the most appropriate parameter values to obtain the most suitable design. Hence, it needs to find the *inverse transformation* from the performance back to the parameters as represented by the "light bulb" in Fig 14. The Y-chart environment is realized by means of Retargetable Simulation. It links parameter values used to define an architecture instance to performance numbers M.4 M.4 M.4.

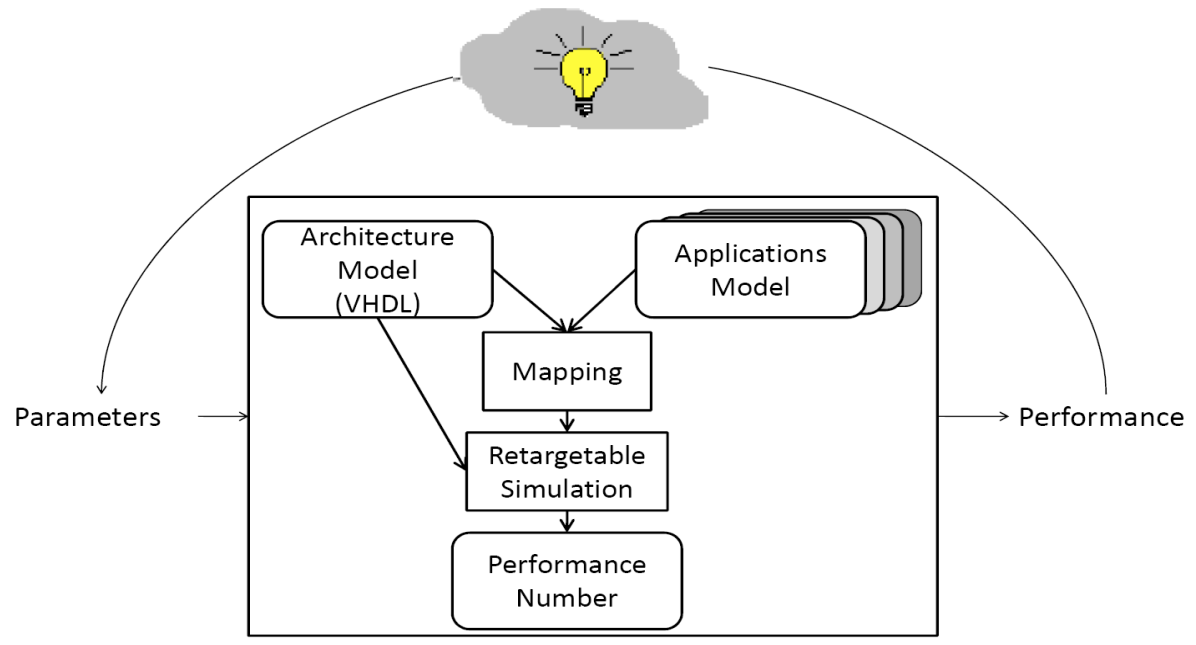

**Fig 14 The *inverse transformation* from the performance back to the parameters.**

The depth of the FIFOs is 32 for both versions. The width for Version 1 is 24-bits, and 8 bits for Version 2. Note that in Version 2; FIFOs do not only exist in the VCs, but also in the NA at the input of NoC (shown in Fig 6). The FPGA is an Altera EP2S1F484C3 implemented with Quartus II 7.2 with DK Design Tool used for the synthesis tool.

## M.1. Results of Version 1 (parameter: depth of the FIFOs/VCs, performance: device utilization)

Fig 15 presents the DSE of the proposed communication architecture. The width of Version 1 FIFOs is 24 bits. The depth of FIFO is the number of packets stored in the FIFO. This version corresponds to the case where all the data have the same lengths.

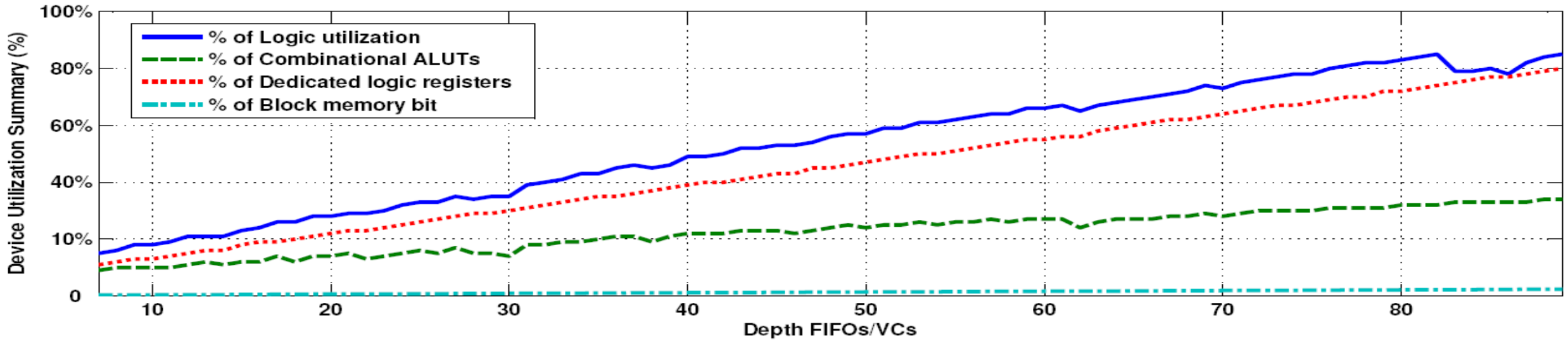

**Fig 15 The Device Utilization Summary of Version 1 on Altera Stratix II. .**

Fig 15 shows that with the increasing of the depths of VCs, the device utilization increases almost linearly. With the maximum depth 89, Logic Array Blocks (LABs) are not enough for the architecture implementation. The memory blocks' augmentation is two times bigger than the augmentation of the total registers.

When DSE reached the maximum depth of FIFOs/VCs, the utilization of ALUTs is 80%, but for the block memory, it has been only used at 5%, that is the synthesis tool does not use memory blocks here as target for implementation of FIFOs.

### M.2. Results of Version 1 (parameter: width of FIFOs/VCs, performance: device utilization)

In Fig 16, the DSE uses two depths for the FIFOs: a 8-data depth (depicted as solid line) and a 32-data depth (presented as dotted line). As data can be parameterized in flits in the NoC, the size of packets is from 20 bits to 44 bits. For the data width, we take from 8-bit data (minimal size for the defined data in the multispectral image analysis architecture) corresponding to 20-bit packets (add 12 bit *header/tail*) to 32-bit data corresponding to 44-bit packets. The X-axes in Fig 16 represents the width of FIFOs/VCs which is the length of packets in the transmission (we consider here size of packet = width of FIFOs).

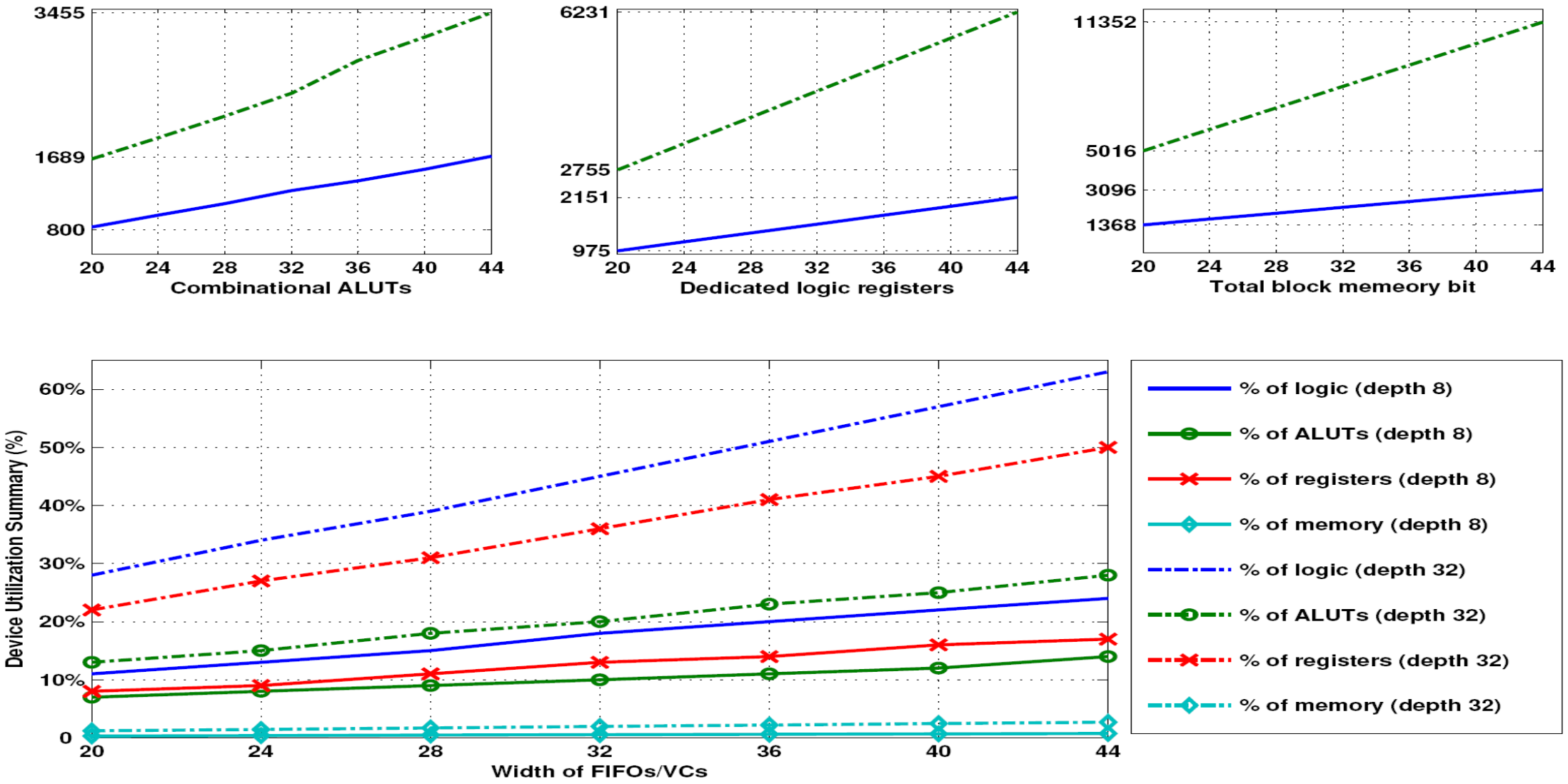

**Fig 16 The device utilization summary with fixed depth of FIFOs/VC but different width on Altera Stratix II for Version 1.**

Results show that the number of resources depends on the width of the FIFOs. The limiting parameters in the size of FIFOs are the number of logic and registers. With a depth of 32, 20% of the registers are used and 40% of the logic is used. The use of logic grows more significantly with a depth of 32. All required resources can be found from a linear equation extracted from the figure and resource predictions can be made without requiring any implementation. We have the same comment on memory blocks.

### M.3. Results of Version 2 (parameter: depth of the FIFOs/VCs, performance: device utilization)

To solve the problem of fixed width of Version 1, Version 2 uses *flits* method. Version 2 has different lengths of the transmitted data which present different widths for each input of the NoC communication. The total data bit transmitted per data is 224 bits (72 bits + 64 bits + 64 bits + 24 bits).

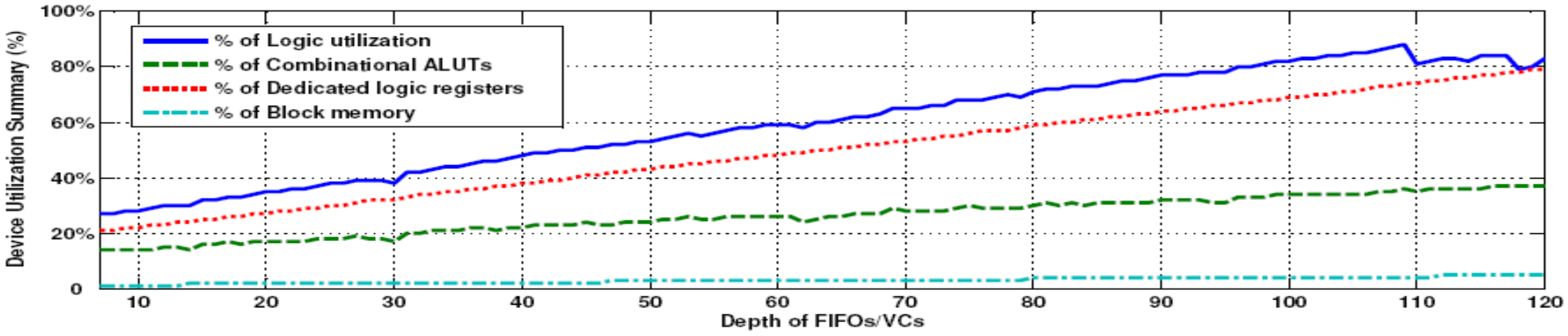

**Fig 17** The Device Utilization Summary on Altera Stratix II for Version 2.

Fig 17 shows the resource utilization summary on Stratix II for Version 2 which has the similar characteristics as the one of Version 1. In the data transmission of Version 2, all the data are divided with 8-bit flits by the NA, which reduce the general width of FIFOs in VCs.

Comparing these 2 versions, Version 1 has fixed widths which is suitable for data having the same size. The structure of Version 1 is simpler, requires fewer resources and has better latency. But for large different lengths/sizes of data transmission, Version 2 is better than Version 1 because it can adapt precisely to the data sizes to obtain an optimal solution.

## M.4. Results of Version 2 (parameter: synthesis tool, performance: device utilization)

Two different synthesis tools have been chosen: DK design Suite V. 5.0 SP5 from Mentor Graphics M.4 and Synopsys Design Compiler M.4 to analyze the impact of synthesis tools on the DSE. From a single VHDL description corresponding to version 2, these two tools gave quite different synthesis results with the same Altera Stratix II EP2S15F484C3 (6240 ALMs) as depicted in Fig 18.

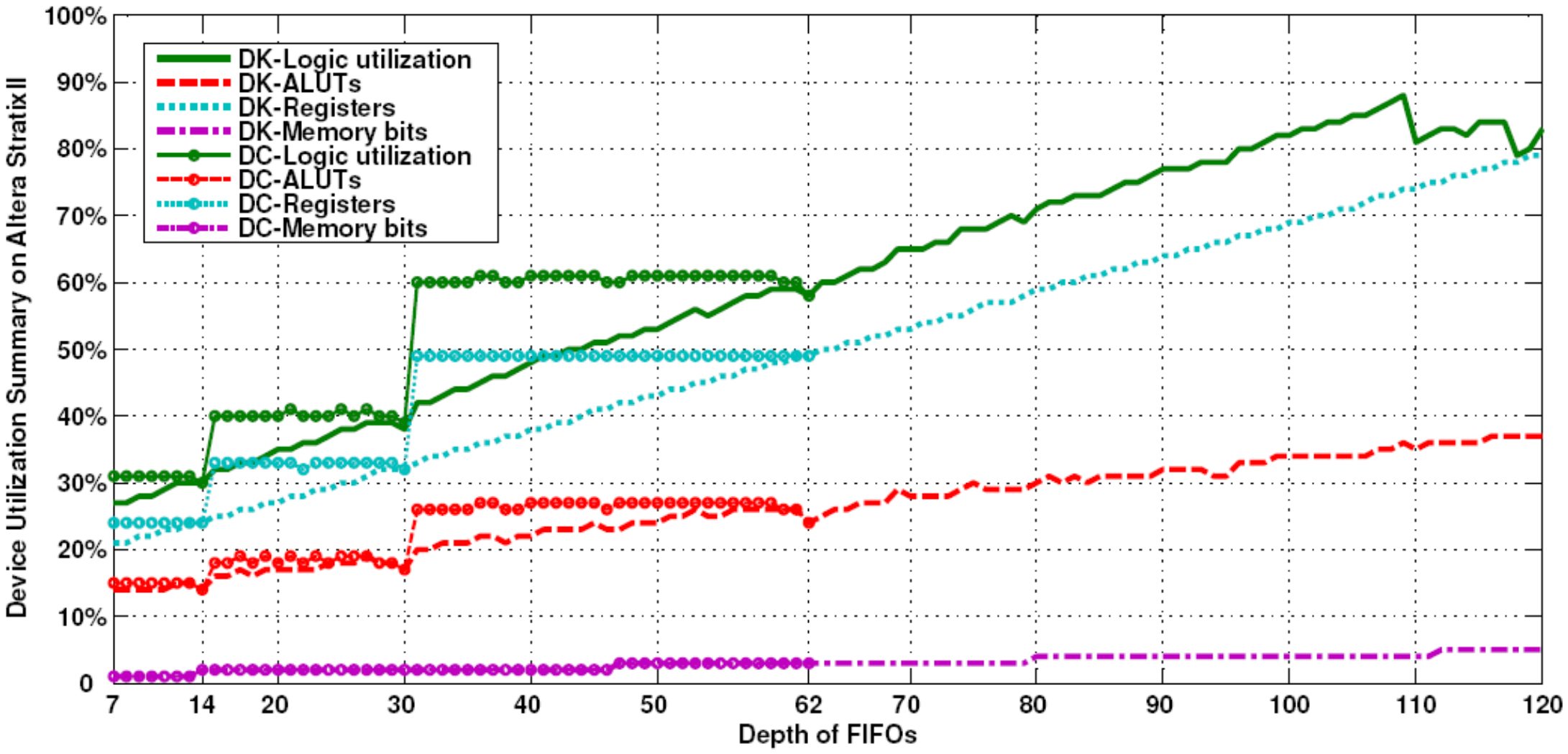

**Fig 18 The Device Utilization Summary on Altera Stratix II for Version 2 with different synthesis tools: Design Compiler and DK design suite.**

These two tools have been chosen as the default synthesis tools on QuartusII 7.1 & 7.2. All the lines with markers ("o", "x", etc) present the synthesized results produced by DK design suite. All the lines without the marker are the results synthesized by Design Compiler. The maximum depths synthesized by these two versions are: 63 by Design Compiler and 120 by DK design suite. For the memory block utilization, these two tools behave in the same way. But for the other implementation factors, results are quite different. At certain depths of the FIFOs/VCs, the resource utilizations are similar (exp.14, 30, 62). However, in most of other depths situations, synthesized results of DK design suite are better than the one of Design Compiler which means it takes fewer resources on the FPGA. The resource utilization increase is quite different as well. With the Design Compiler synthesis tool, device utilization increases as several "steps" but with DK design suite, it increases more smoothly and linearly. In this case, the synthesis with DK design suite makes resource utilization prediction much easier than with Design Compiler.

## Conclusion and Perspectives

The presented architecture is a parameterized architecture dedicated to image analysis applications on FPGA. All flows and data are analyzed to propose two generic NoC architectures, a ring for results and command and a dedicated FT NoC for data. Using both communication architectures, the designer inserts several modules; the number and type of modules depend on the algorithm requirements. The proposed NoC for data transfer is more precisely a parameterized TDM architecture which is fast, flexible and adaptable to the type and size of data used by the given image analysis application. This NoC uses a Fat Tree topology with VC packet-switching and parameterized flits.

According to the implementation constraints, area and speed, the designer chooses one version and can adapt the communication to optimize the area/bandwidth/latency tradeoff. Adaptation consists in adding several switches in parallel or in serial and to size data (and flit), FIFOs and Virtual channels for each switch. Without any implementation the designer can predict the resources used and required. This Fat Tree generic topology allows us to generate and explore systematically a communication infrastructure in order to design efficiently any dataflow image analysis application.

Future work will focus on automating the exploration of the complete architecture and the analysing of the algorithm architecture matching according to the different required data. From the evaluation of the NoC exploration an automated tool can predict the most appropriate communication architecture for data transfer and the required resources. The power analysis will be analyzed to complete the Design Space Exploration of the NoC architecture. Power, area and latency/bandwidth are the values which will guide the exploration process.

## References

[1] Paul Taylor, "Nvidia opens mobile GPU kimono: Slideware shows higher performance, lower TDPs," electronic resource available: http://www.theinquirer.net/inquirer/news/1271809/nvidia-mobile-gpu-kimono, Wednesday, 17 June 2009.

[2] Zhang Yuhong and al., "A system verification environment for mixed-signal SOC design based on IP bus," *ASIC, 2003. Proceedings. 5th International Conference on*, vol. 1, 2003, 278-281 Vol.1.

[3] U. Farooq, M. Saleem, and H. Jamal, "Parameterized FIR Filtering IP Cores for Reusable SoC Design," *Information Technology: New Generations, 2006. ITNG 2006. Third International Conference on*, 2006, 554-559.

[4] Soo Ho Chang and Soo Dong Kim, "Reuse-based Methodology in Developing System-on-Chip (SoC)," *Software Engineering Research, Management and Applications, 2006. Fourth International Conference on*, 2006, 125-131.

[5] F. Moraes and al., "HERMES: an infrastructure for low area overhead packet-switching networks on chip," *Integration, the VLSI Journal* 38, no. 1 (2004): 69-93.

[6] Théodore Marescaux and al., "Interconnection Networks Enable Fine-Grain Dynamic Multi-tasking on FPGAs," *Field-Programmable Logic and Applications: Reconfigurable Computing Is Going Mainstream*, 2002, 741-763, http://dx.doi.org/10.1007/3-540-46117-5_82.

[7] C. A. Zeferino and A. A. Susin, "SoCIN: a parametric and scalable network-on-chip," *Integrated Circuits and Systems Design, 2003. SBCCI 2003. Proceedings. 16th Symposium on*, 2003, 169-174.

[8] E. Salminen, A. Kulmala, and T.D. Hamalainen, "HIBI-based multiprocessor SoC on FPGA," *Circuits and Systems, 2005. ISCAS 2005. IEEE International Symposium on*, 2005, 3351-3354 Vol. 4.

[9] C. Hilton and B. Nelson, "PNoC: a flexible circuit-switched NoC for FPGA-based systems," *IEE Proceedings-Computers and Digital Techniques* 153, no. 3 (2006): 181-188.

[10] C. Bobda and A. Ahmadinia, "Dynamic interconnection of reconfigurable modules on reconfigurable devices," *Design & Test of Computers, IEEE* 22, no. 5 (2005): 443-451.

[11] Hyung Gyu Lee and al., "Design space exploration and prototyping for on-chip multimedia applications," *Design Automation Conference, 2006 43rd ACM/IEEE*, 2006, 137-142.

[12] Ahmadinia and al., "A practical approach for circuit routing on dynamic reconfigurable devices," *Rapid System Prototyping, 2005. (RSP 2005). The 16th IEEE International Workshop on*, 2005, 84-90.

[13] Fresse, A. Aubert, N. Bochard "A Predictive NoC Architecture for Vision Systems Dedicated to Image Analysis," EURASIP Journal on Embedded Systems Volume, Article ID 97929, 13 pages, 2007.

[14] G.Schelle, D. Grunwald, "Exploring FPGA network on chip implementations across various application and network loads," in the proceeding of FPL08, International Conference on Field Programmable Logic and Applications, pp.41-46, Sep 8-10, 2008.

[15] P. Wagener, "Metastability-a designer's viewpoint," dans *ASIC Seminar and Exhibit, 1990. Proceedings., Third Annual IEEE*, 1990, P14/7.1-P14/7.5.

[16] E. Brunvand, "Implementing Self-Timed Systems with FPGAs," FPGAs, W. Moore and W. Luk, eds., Abingdon EE&CS Books, Abingdon, England, 1991, pp. 312-323.

[17] W.J. Dally, "Virtual-Channel flow control," in *IEEE* Parallel and Distributed Systems; vol.3, Issue.2, pp.194-205, 1992.

[18] E. Rijpkema, K. G. W. Goossens, A. Rădulescu, J. Dielissen, J. Van Meerbergen, P. Wielage, E. Waterl, "Trade offs in the design of a router with both guaranteed and best-effort services for networks on chip," in *IEEE* Proc. Computers and Digital Techniques, Vol.150, Issue.5, pp.294-302, 2003.

[19] H.S. Wang, L.-S. Peh, S. Malik, "A power model for routers: Modeling Alpha 21364 and InfiniBand Routers," in Proc 10th High Performance Interconnects 2002, pp.21-27, 2002.

[20] P. Gupta, N. McKeown, "Designing and implementing of a Fast crossbar scheduler," in *IEEE* Micro Jan/Feb 1999, vol.19, Issue.1, pp.20-28,1999.

[21] W.J. Dally, B. Towles "Route packets, not wires: on-Chip interconnection Networks," in Proc. DAC'01, Design Automation Conference, pp.684-689, 2001.

[22] F. Koning, W, Praefcke, "Multispectral image encoding," in Proceeding of ICIP 99, International Conference on Image Processing, Vol.3, pp.45-49, Oct.24-28, 1999.

[23] Kaarna, P. Zdmcik, H. Kalviainen, J Parkkinen, "Multispectral image compression," in Proceeding of Fourteenth International Conference on Pattern Recognition, Vol. 2, pp.1264-126, Aug.16-20, 1998.

[24] D. Tretter, C.A. Bouman, "Optimal transforms for multispectral and multilayer image coding," in IEEE Image Processing, Vol.4, issue 3, pp. 296-308, March 1995.

[25] P.Zemcik, M.Frydrych, H. Kalviainen, P.Toivanen, J. Voracek, " Multispectral image colour encoding," in Proceeding of 15 th international conference on Pattern Recognition, Vol.3, pp. 605-608, Sep. 3-7, 2000.

[26] Manduca, "Multispectral image visualization with nonlinear projections," in IEEE Image Processing, vol.5, issue 10, pp. 1486-1490, Oct.1996.

[27] D. Tzeng, "Spectral-based color separation algorithm development for multiple-ink color reproduction," Ph.D. Dssertation, R.I.T., Rochester, NY, 1999.

[28] E.A. Day, "The Effects of Multi-channel Spectrum Imaging on Perceived Spatial Image Quality and Color Reproduction Accuracy," M.S. Thesis, R.I.T., Rochester, NY, 2003.

[29] L. Zhang, A.-C. Legrand, V. Fresse, V. Fischer, "Adaptive FPGA NoC-based Architecture for Multispectral Image Correlation," in Proc. IS&T CGIV&MCS08, IS&T's fourth European Conference on Colour in Graphics, Imaging, and Vision, and MCS'2008, the 10th International Symposium on Multispectral Colour Science, pp.451-456. Terrassa, Barcelona, Spain, June 9-13,2008.

[30] A.C.J. Kienhuis, Ir.E.F. Deprettere, "Design Space Exploration of Stream-based Dataflow Architectures: Methods and Tools," Thesis report, Toegevoegd Promotor, Technische Universität Braunschweig, 1999.

[31] H.P. Peixoto, M.F. Jacome, "Algorithm and architecture-level design space exploration using hierarchical data flows," in proceeding of IEEE international Application-Specific Systems, Architectres and Processors, pp.272-282, July 14-16, 1997.

[32] V. Krishnan, S. Katkoori, "A genetic algorithm for the design space exploration of datapaths during high-level synthesis," in IEEE Evolutionary Computation, vol.10, issue.3, pp.213-229; June 2006.

[33] Xilinx Virtex4, http://www.xilinx.com/products/silicon_solutions/fpgas/virtex/virtexS/index.htm.

[34] ALM structure of Stratix II, white paper "Stratix II vs. Virtex-4 Performance Comparison," available: http://www.altera.com/literature/wp/wp_s2v4perf.pdf.

[35] Mentor graphics "DK Design Suite Tool", http://www.agilityds.com/products/c_based_products/dk_design_suite/

[36] RTL-to-Gates Synthesis using Synopsys Design Compiler, available: http://csg.csail.mit.edu/6.375/6_375_2008_www/handouts/tutorials/tut4-dc.pdf.